\documentclass{JHEP3}
\usepackage{epsfig,amsmath,amsthm,latexsym}
\newcommand{\be}{\begin{equation}}
\newcommand{\ee}{\end{equation}}
\newcommand{\bea}{\begin{eqnarray}}
\newcommand{\eea}{\end{eqnarray}}

\def\IZ{\relax\ifmmode\hbox{Z\kern-.4em Z}\else{Z\kern-.4em Z}\fi}
\newcommand{\IS}{{\bf S}}

\newcommand{\non}{\nonumber \\}

\def\half{{\frac12}} 

\def\del{{\partial}}

\def\cl{{\cal L}}

\def\al{\alpha} \def\bt{\beta}
\def\gm{\gamma}  

\def\hal{{\hat \alpha}} \def\hbt{{\hat \beta}}
\def\hgm{{\hat \gamma}} 
 
  \def\tphi{\tilde{\phi}}

\def\presub{\vspace{.5cm} \noindent}

\def\bi{\begin{itemize}} \def\ei{\end{itemize}}

\def\Schw{Schwarzschild }
\def\({\left(} \def\){\right)}
\def\[{\left[} \def\]{\right]}

\newcommand{\Om}{\Omega}
\newcommand{\om}{\omega}
\newcommand{\N}{\nabla}
\newcommand{\Nt}{\tilde{\nabla}}

\newcommand{\lm}{\lambda}
\newcommand{\Gm}{\Gamma}
\newcommand{\hb}{\hbar}
\newcommand{\tl}{\tilde}
\newcommand{\Dl}{\Delta}
\newcommand{\Ch}{\hat C_2}

\preprint{{\tt hep-th/0703283}}

\title{ \center{Dynamical vs. Auxiliary Fields in Gravitational Waves around a Black Hole}}

\author{Vadim Asnin and  Barak Kol\\
 Racah Institute of Physics\\
 Hebrew University\\
 Jerusalem 91904, Israel\\
E-mail: \email{vadima@pob.huji.ac.il},
{\tt\href{mailto:barak_kol@phys.huji.ac.il}{barak\_kol@phys.huji.ac.il}}}

\abstract{The auxiliary/dynamic decoupling method of hep-th/0609001
applies to perturbations of any co-homogeneity 1 background (such as
a spherically symmetric space-time or a homogeneous cosmology). Here
it is applied to compute the perturbations around a Schwarzschild
black hole in an arbitrary dimension. The method provides a clear
insight for the existence of master equations. The computation is
straightforward, coincides with previous results of Regge-Wheeler,
Zerilli and Kodama-Ishibashi but does not require any ingenuity in
either the definition of variables or in fixing the gauge. We note
that the method's emergent master fields are canonically conjugate
to the standard ones. In addition, our action approach yields the
auxiliary sectors.}

\begin{document}

\section{Introduction and Summary}

Recently an improved understanding of perturbations around
backgrounds with (at most) one non-homogeneous coordinate
(``co-homogeneity 1'') was achieved \cite{1dPert}. Two cases of
physical interest fall under the title of ``co-homogeneity 1
backgrounds'': static spherically-symmetric space-times and
homogeneous cosmologies. In this paper we apply the method to the
\Schw black hole in an arbitrary dimension, re-deriving the master
equations for the perturbation spectrum and wave profiles while
obtaining new insights to be mentioned below.

The method of \cite{1dPert} includes the method known as ``gauge
invariant perturbation theory'' together with an explicit treatment
of the auxiliary sector hence we refer to it as the
``\emph{auxiliary/dynamic decoupling method}''. In addition it was
proven in  greater generality, demonstrating that the essential
condition is the dimension of co-homogeneity (1d).

Before discussing the specifics of black hole perturbations let us
recall the basic ingredients of the general theorem \cite{1dPert}.
By performing dimensional reduction over the homogeneous coordinates
we may consider the problem to be essentially 1d. It is found that a
1d action with $n_F$ fields invariant under $n_G$ gauge functions
can be decoupled into two sectors, auxiliary and dynamic, where the
auxiliary sector is purely algebraic as its name suggests and all
fields in both sectors are gauge invariant. The auxiliary sector
consists of $n_G$ fields which are nothing but the (shifted) metric
and vector fields from the 1d perspective, and hence both have
formally minus one degrees of freedom. The dynamic sector contains
$n_F-2\, n_G$ fields which remain after the gauge functions are
eliminated from the $n_F-n_G$ non-auxiliary fields. All field
redefinitions are local in 1d and invertible.

In addition to the decoupling between auxiliary and dynamic fields
there is another useful decoupling mechanism. From the 1d
perspective the isometry group becomes the theory's gauge group
under which the fields are charged. Therefore, the quadratic action
decouples according to representations of this group.

Chronologically, the general theorem of \cite{1dPert} followed by
generalizing ``the'' derivation of the Gregory-Laflamme mode
\cite{power} (see \cite{rev,HOrev} for reviews of the black-hole
black-string transition as well as \cite{TopChange}).

\vspace{0.5cm}

Let us return to discuss the specific case of perturbations around
Schwarzschild. Of course, the results for this case were already
known. The 4d case is by now classic: the odd (vector) sector
reduces to the Regge-Wheeler master equation \cite{RW} while the
even (scalar) sector reduces to the somewhat more cumbersome Zerilli
master equation \cite{Zerilli}. The higher dimensional
generalization however, is relatively recent, and was achieved by
Kodama and Ishibashi \cite{IshibashiKodama} (see also
\cite{dialogue1} for the master equation of static perturbations
through a different method, and \cite{IshibashiKodama2} for an
application to a stability analysis of higher dimensional \Schw ).

Our method reproduces and confirms these results while adding the
form of the auxiliary sectors and providing the advantages of a
different point of view (somewhat in analogy to Moncrief's method
\cite{Moncrief1974} in 4d, see also
\cite{MartelPoisson,NagarRezzolla} for some recent discussions of
perturbations of the \Schw black hole in the gauge-invariant
formalism). The method has the advantage of being practically
straightforward. We use an action approach and simplify it in a
natural way. We explain why the equations in each sector can be
reduced to a single master equation (as anticipated in
\cite{1dPert}). Moreover, there is no need for ingenious field
redefinitions as in \cite{IshibashiKodama} nor is there a need for
choosing a gauge-fix as in \cite{RW}. In addition our treatment is
independent of the choice of coordinates for the background \Schw
space-time.

While our method identifies the space of dynamic fields, it only
fixes the definition of the dynamic fields up to a
linear\footnote{To preserve the linearity of the equations of
motion.}, radially-dependent canonical transformation. It is found
that the action is considerably simpler for some choices, and the
standard master fields of Regge-Wheeler, Zerilli and
Kodama-Ishibashi are essentially canonically conjugate to the ones
which naturally emerge from our method. The standard master fields
have the additional property that their $r$ and $t$ derivatives
combine in the master equation to the form of a Laplacian in the
reduced 2d $(r,t)$ space. This property comes as a surprise from the
current 1d perspective which concentrates on the $r$ coordinate, and
to gain insight into it, it is probably necessary to study the
properties of perturbations on co-homogeneity 2 spaces. Another open
question from our perspective is whether the standard form of the
equations is in some sense optimal or whether further simplification
is possible, especially in the scalar potential which is quite
involved.

We start in section \ref{review} with a review of the
auxiliary/dynamic decoupling method. In section
\ref{Lagrangian-section} we derive the quadratic Lagrangian for the
perturbations. While the derivation of this Lagrangian is completely
straightforward, it is quite lengthy, and perhaps there is room to
make it more economical. Due to the decoupling of representations at
quadratic order (mentioned above) the Lagrangian decouples to three
sectors known as tensor, vector and scalar, and which reduce in the
4d case to the even and odd sectors (the tensor sector degenerates).

In section \ref{decouple-section} we proceed to apply the method to
this Lagrangian. The tensor sector has one field and no gauge
functions ($n_F=1 ~n_G=0$) and hence requires no treatment. The
vector sector has $n_F=3 ~n_G=1$ and since $3-2*1=1$ we immediately
expect a single dynamic field,  and hence a decoupled equation of
motion, namely a master equation. Indeed our straightforward
manipulation results in the dynamic and auxiliary sectors of the
action. The dynamic sector encodes a master equation which is shown
to be equivalent to the Kodama-Ishibashi equation (or Regge-Wheeler
in 4d). The scalar sector is more involved: it has $n_F=7 ~n_G=3$
but as $7-2*3=1$ there is still a single master equation. Again we
follow our method to derive the auxiliary sector together with a
master equation and the latter is confirmed to be equivalent to the
Kodama-Ishibashi equation (Zerilli's in 4d). Our main results are
summarized in section \ref{summary}.

It would be interesting to apply the auxiliary/dynamic decoupling
method to other space-times including perturbations around the
critical merger solution \cite{AKS}.

\section{Review of the auxiliary/dynamic decoupling method}
\label{review}
Consider perturbations around a space-time $X$ which
is of co-homogeneity 1, namely that all but one of its coordinates
(denoted by $r$) \footnote{In \cite{1dPert} the essential coordinate
was denoted by $x$, but here since we specialize to the spherically
symmetric case we denote it by $r$. Some additional notation which
differs between this paper and \cite{1dPert} is:
 ``DG'' which stood for ``derivatively-coupled''
sector was exchanged for ``A'' which stands for ``auxiliary'' or
``algebraic'', and ``X'' which stood for the dynamic sector was
traded with ``Dyn''.}
 belong to a homogeneous space. One starts by
effectively reducing the problem to a 1d problem on $r$ as
follows. The metric perturbation tensor $h_{\mu\nu}$ decomposes
into several fields according to whether $\mu$ is $r$ or belongs
to the homogeneous space. Furthermore, each component of
$h_{\mu\nu}$ can be expanded into harmonic functions of the
homogeneous space in order to achieve a separation of the
variables. For instance, for the time translation symmetry
$U(1)_t$ we expand in $\exp(i \omega\, t)$, while for the
spherical symmetry $SO(D-1)_\Omega$ we expand in spherical
harmonics (of the appropriate tensor type).

Altogether the action reduces to a 1d action in $r$, with a
certain field content \be
 \phi=\phi^i(r) ~~~ i=1,\dots,n_F \ee
where $n_F$ is the number of (real) fields. Actually the field
content necessarily includes the 1d metric $h_{rr}$ as well as
massless vectors which originate from the isometry group of $X$ and
become the gauge symmetry in 1d. Hence the dimensionally reduced
theory is actually a \emph{1d gauged-gravity theory} with additional
matter content (which generically includes $r$-scalars as well as
massive $r$-vectors and $r$-tensors).

The invariance of $X$ under infinitesimal diffeomorphisms
 $\xi^\mu$ translates into invariance of the 1d action under certain gauge
functions, whose number we denote by $n_G$, as follows \bea
 \delta \phi &=&  G_1\, \xi' + G_0\, \xi \non
 \xi &=& \xi^a(r) \qquad a=1,\dots,n_G \label{def-xi-G} \eea
 where $\(G_1(r),\,G_0(r) \)$ are a pair $n_F \times n_G$ $r$-dependent
matrices.

The action is quadratic in $\phi$ since we are interested only in
linear perturbations, and we furthermore assume it to be quadratic
in derivatives as is the case for classical GR.

The standard counting of equations proceeds as follows. One uses the
$n_G$ gauge functions to set $n_G$ gauge-fixings relations over the
fields, possibly eliminating $n_G$ fields. The gauge-fixing is
accompanied by $n_G$ constraint equations which are usually of lower
order in derivatives as well as $n_f-n_G$ other equations.
\cite{1dPert} proved that under generic conditions \footnote{Namely,
that $G_1$ is non-degenerate, as well as $G_X$ and $L_{DG} \equiv
L_A$ which are respectively the gauge transformation when restricted
to the non-auxiliary sector and the (algebraic) action of the
auxiliary sector, both as defined in \cite{1dPert}.} it is always
possible in 1d to completely eliminate the gauge through a
transformation of the fields which is local (in $r$) and invertible,
and that the action decouples into an \emph{auxiliary sector}
consisting of $n_G$ fields with an algebraic action and a
\emph{dynamic sector} consisting of \be
 n_F-2\, n_G \label{n-dyn} \ee
 fields (generically second order in derivatives). Comparing to
the standard counting we find that \emph{in 1d it is always possible
to have algebraic equations for both the $n_G$ gauge-fixing
relations as well for $n_G$ of the equations of motion}.

As the full proof is given in \cite{1dPert}, we choose to limit
ourselves to a description of the key mechanisms at work. The
essential idea is a sort of \emph{``derivative counting''} applied
to the equation which states that the equations of motion (EOM) are
satisfied for a pure-gauge field configuration. We define
``derivatively gauged'' fields or ``A-fields''\footnotemark[2] to be
the image of $G_1$ (\ref{def-xi-G}). Physically, these are nothing
but the $r$-tensors and $r$-vectors. The term with the highest
number of derivatives (3) originates from the second-derivative
terms in the EOM applied to the A-fields. Since this is the only
term at that order we find that it must vanish, namely that the
A-fields do not appear in action terms with two derivatives.
Moreover, by considering a similar expansion of the second variation
of the action we find that action terms in which only A-field appear
are purely algebraic thus justifying the name ``auxiliary sector''.
The relevant algebraic matrix is denoted by $L_A$. In the generic
case that $L_A$ is non-degenerate we may ``complete the square''
thereby decoupling the A-fields from the rest of the action after a
suitable shift in their definition.

Then we inspect the action of the gauge transformation on the new
fields. The algebraic nature of the action in the auxiliary sector
means that no symmetry is possible in that sector and hence no gauge
invariance. By definition, the non-auxiliary fields are gauged only
algebraically ($\xi'$ does not appear in $\delta \phi$) which means
that the gauge can be completely eliminated by an algebraic
re-definition of fields. Thereby $n_F-2\, n_G$ fields are present in
the other, dynamic sector, and completing the general description of
the method.

\section{The 1d quadratic action}
\label{Lagrangian-section}

We consider gravitational waves in the Schwarzschild background in
an arbitrary space-time dimension $D$. The isometry group of this
space-time is $O(D-1)_\Om \times U(1)_t \times \IZ_{2,T}$, where
$O(D-1)_\Om$ is the spherical symmetry, $U(1)_t$ is time
translation and $\IZ_{2,T}$ is time reversal. These isometries
allow the use of the following ``maximally general ansatz'',
namely an ansatz such that all of Einstein's equations can be
derived from the variation of the gravitational action with
respect to its fields
 \be
 ds_0^2=e^{2a(r)}\, dt^2+e^{2b(r)}\, dr^2 + e^{2c(r)}\, d\Om_{D-2}^2 ~,
  \label{background}\ee
where $d\Om_{D-2}^2$ is the standard metric on the $D-2$ sphere. The
standard \Schw coordinates are \be
 e^{2 a}=f(r) \equiv 1-(r_0/r)^{D-3}~,\qquad e^{2 b}=f^{-1},\qquad e^{2 c}=r^2 ~, \label{Schw-coord} \ee
 but we may consider other coordinate systems
(namely the choice of $r$) such as in \cite{nSchw},
  and for generality we shall not restrict to any specific choice.
A general transformation from \Schw coordinates shows that for any
choice of coordinate we have \be c'=e^{a+b-c}~,\qquad
a'=-(D-3)e^{b-c}\sinh a ~,\ee
 where prime means a derivative w.r.t. $r$.

The metric in the gravitational wave is
$g_{\mu\nu}=g^{(0)}_{\mu\nu}+h_{\mu\nu}$, where $g^{(0)}_{\mu\nu}$
stands for the background metric and $h_{\mu\nu}$ is a perturbation.
In order to write down the action that describes gravitational waves
in a given background to linear order it is convenient to define the
following tensors: \be\hb_{\mu\nu}:=h_{\mu\nu}-\frac12
g_{\mu\nu}h_{\lm}^{\lm}\label{hbar}~,\qquad\hbar:=\hbar^{\mu}_{\mu}~,\ee
and
\be\gm^\lm_{~\mu\nu}:=\frac12(\N_{\mu}\hb_{\nu}^{\lm}+\N_{\nu}\hb_{\mu}^{\lm}-\N^{\lm}\hb_{\mu\nu})~,\ee
which is similar to the linearized part of the Christoffel symbol
only with $h \to \hbar$. Using these definitions the quadratic
action can be compactly written as \be
 S=\int d^Dx \sqrt{g}\, \bigl(2\, \gm^\lm_{~\mu\nu}\, \gm_{~~\lm}^{\mu\nu}+\frac{1}{2(D-2)}\N_{\mu}\hb
\N^{\mu}\hb\bigr)\label{MultiDAction} ~.\ee

The gauge invariance of the full Einstein-Hilbert action has two
consequences for the quadratic action: it is invariant both under a
change of background coordinates (since it is manifestly invariant
under local Lorentz transformations) and under the following gauge
symmetry acting on the fields $h_{\mu\nu}$ \be
 \delta h_{\mu\nu} = \N_{\mu}\xi_{\nu} + \N_{\nu}\xi_{\mu}
 \label{TotalGauge},\ee
where $\xi_{\mu}$ is an arbitrary vector field. These two
symmetries can be understood in a following way: as the components
of the metric in the Einstein-Hilbert action are expanded around
the background, also the gauge functions should be expanded; an
invariance under reparameterizations appears in the zeroth order
and the invariance (\ref{TotalGauge}) appears in the first order.

In this section we outline the derivation of an action in the radial
direction, via a dimensional reduction of (\ref{MultiDAction}), and
the details of the derivation are presented in Appendix
\ref{1DAction}. We distinguish the components of the metric
perturbation $h_{\mu\nu}$ according to their tensor type, both with
respect to $r$ and with respect to the angular coordinates. Namely,
in the radial direction $h_{rr}$ is an $r$-tensor, $h_{rt}$ and
$h_{ri}$ (we denote by Latin indices the angular coordinates) are
$r$-vectors, and the rest are $r$-scalars. Similarly on the sphere
$h_{tt}$, $h_{tr}$ and $h_{rr}$ are $\Om$-scalar-fields, $h_{ti}$
and $h_{ri}$ are $\Om$-vectors-fields and $h_{ij}$ form an
$\Om$-tensor-field.

As usual during dimensional reduction we expand each field into
harmonic functions with respect to the isometry group. Actually the
isometry group becomes the gauge group from the 1d (radial) point of
view, and each field is labeled by its representation under this
group. For the spherical coordinates we need the scalar, vector and
symmetric tensor spherical harmonics. The properties of these
harmonics and the conventions used here are discussed in appendix
\ref{SphericalHarmonics}. As discussed there, a spherical harmonic
of a given type is labeled by a non-negative integer number $l$ that
determines the Dynkin indices $[l_1,...l_n]$ of the irreducible
representation of $SO(D-1)$, and individual functions within the
representation are further
 labeled by $m_1,...m_n$ which we denote collectively by $m$.
 There is a single family of scalar spherical harmonics, namely $Y^S_{l m}$
belonging to the representation $[l,0...0]$, two families of
vector spherical harmonics, namely $Y^S_{i,\ l m}$ in the
representation $[l,0...0]$ and $Y^V_{i,\ l m}$ in $[l,1,0...0]$,
and 4 families of tensor spherical harmonics: $Y^S_{ij,\ l m}$ and
$\tl Y^S_{ij,\ l m }$ in $[l,0...0]$, $Y^V_{ij,\ l m}$ in
$[l,1,0...0]$ and $Y^T_{ij,\ l m}$ in $[l,2,0...0]$. The
superscripts $S,\,V,$ and $T$ stand respectively for scalar,
vector and tensor and denote the three relevant families of irreps
$[l,0...0],\, [l,1,0...0]$ and $[l,2,0...0]$, according to the
lowest spin field for which the specific irrep family appears.
Thus we should distinguish between two angular tensor types. The
first tensor-type is with respect to local diffeomorpshims under
$GL(D-1)$ is denoted, as was mentioned above, by  $\Om$-scalar
-field, $\Om$-vector-field etc. The second tensor type is with
respect to the $SO(D-1)$ global isometries and will be denoted by
$S,\,V$ and $T$ as above. From hereon we shall mostly refer to the
latter tensor type (which specifies the representations of fields
under the gauge group from the 1d perspective), and when there is
a possibility of confusion we shall call such types
$SO(D-1)$-scalars, vectors or tensors.

Dimensional reduction with respect to $t$ means a Fourier expansion.
Altogether the expansion is
\begin{gather}
h_{tt}=\int d\om\, \sum\limits_{l,m_1...m_n}\, h^S_{tt,l,m_1...m_n}\, Y^S_{l,m_1...m_n}\, e^{i\om t},\nonumber\\
h_{tr}=h^S_{tr}\, Y^Se^{i\om t}~,\qquad h_{rr} = h^S_{rr}\, Y^Se^{i\om t}~,\nonumber\\
h_{ti}=\(h^S_{t}\, Y^S_i+h^V_{t}\, Y^V_i \)\, e^{i\om t}~,\qquad h_{ri}=\(h^S_{r}\, Y^S_i+h^V_{t}\, Y^V_i\) e^{i\om t}~,\nonumber\\
h_{ij}=(h^S\, Y^S_{ij}+\tl h^S\tl Y^S_{ij} + h^V\, Y^V_{ij}+h^T\,
Y^T_{ij})\,e^{i\om t}\label{Expansions}~.
\end{gather}
 where only the first equation shows explicitly the summation over
representations $l$, over the indices $m \equiv m_1...m_n$ and the
$\om$ integration, and it should be understood to be implicit in
all the other equations as well as those to follow.

The notation for the coefficients here is the following: The
superscripts denote the $SO(D-1)$ tensor-type (or irrep family) of
the corresponding spherical harmonic, and the subscripts give the
non-angular indices of $h_{\mu\nu}$.

The gauge variations of $h_{\mu\nu}$ (\ref{TotalGauge}) allow one to
calculate the gauge variations of the coefficients in these
expansions if one expands also the components of the vector
$\xi_{\mu}$ into the spherical harmonics and Fourier expands in
time: \be
 \xi_t=\xi^S_t\, Y^S\, e^{i\om t},\qquad
 \xi_r=\xi^S_r\, Y^Se^{i\om t},\qquad \xi_i=(\xi^S\, Y^S_i+\xi^V\, Y^V_i)\, e^{i\om t},
\label{GaugeComponents}\ee where the notation is similar to
(\ref{Expansions}). We shall spell out the gauge variation of the
fields after dimensional reduction in the next section.

The next step is to plug the expansions (\ref{Expansions}) into
the action and calculate the angular integrals (all necessary
formulae are given in Appendix \ref{SphericalHarmonics}). In the
next section we present the resulting action sector by sector in
(\ref{action-sectors}, \ref{TL}, \ref{VectorLagrangian},
\ref{ScalarLagrangian} ) and proceed to further analyze and
decouple it.

\subsection*{4d}

Our above-mentioned considerations do not apply directly to 4d
since the rank of $O(3)$, the rotational symmetry group, is
smaller than 2 and the above field-theoretic definitions of the
vector and tensor families become meaningless since they are
defined through the Dynkin indices $[l,1,0 \dots 0]$ and $[l,2,0
\dots 0]$. However, since the decoupling into sectors relies
mostly on the orthogonality relations of the spherical harmonics,
the decoupling continues to hold in 4d, though with some changes:
\emph{the scalar sector becomes the even sector, the vector sector
becomes the odd sector and the tensor sector disappears}, as we
now proceed explain. Moreover, we find that the final form of the
decoupled action in 4d can be obtained simply by setting $D=4$ in
the expressions we have for an arbitrary $D$.

One change is that in 4d there is no tensor sector. To understand
that we note that on $\IS^2$ the metric has only 3 components, and
therefore the general decomposition of the metric into 4 components
has to degenerate, and it turns out that the tensor component is the
one that vanishes identically.

The scalar and vector sectors still exist, but their
field-theoretic definition is different. The group of isometries
$O(3)$ includes also a $\IZ_2$ of reflections in its center.
Therefore any representation can be classified as either even or
odd, and the quadratic action must decouple into these two
sectors. Actually we assign a parity to each field, according to
the parity of its $l=0$ component, and for higher $l$ the parity
gets multiplied by $(-)^l$. In this way the fields $h_{tt}^S$,
$h_{tr}^S$, $h_{rr}^S$, $h_t^S$, $h_r^S$, $h^S$ and $\tilde h^S$
are found to have even parity, whereas $h_t^V$, $h_r^V$ and $h^V$
are odd. Equivalently, the scalar sector becomes the even sector
in 4d and the vector sector becomes the odd sector.

The case of $5d$ is also special from the group theoretical point of
view since $SO(4)$ decomposes into $SU(2) \times SU(2)$. However, in
this case no sector degenerates, and our general theory continues to
hold, since the rank is 2, and due to the reliance on the
orthogonality relations.

\section{Decoupling each sector}
\label{decouple-section}

The quadratic action decouples into three parts: scalar, vector and
tensor \be
 S = S^S + S^V + S^T ~, \label{action-sectors} \ee
due to the orthogonality property of the spherical harmonics. In
this section we shall further decouple each one of the sectors
separately.

In each sector the general formula for counting dynamic fields
(\ref{n-dyn}) leads us to expect a single dynamic field, which is
called ``the master field''. This master field is unique up to an
$r$-dependent factor, which we attempt to choose so as to simplify
the action, but we do not pretend to optimize this choice here.
Moreover, one is allowed to use canonical transformations which
are non-Lagrangian (namely, the new field is a function not only
of the old field, but also of its momentum). And indeed, we find
that in order to obtain the standard form of the master equations
we need to transform to a new field which is essentially the
momentum conjugate to our original field.

Another general feature is the $r$-tensor-type. All master fields
are $r$-scalars, while the auxiliary fields are $r$-vectors and
$r$-tensors. As usual, local Lorentz invariance in 1d determines the
required power of $g^{rr}=e^{-2 b}$ for each term, so we should be
able to present the action such that $b$ does not show up
explicitly. To this end we use the following standard notation for
derivatives and the integration measure \be
 \del^r \equiv g^{rr}\, \del_r= e^{-2b}\, \del_r \qquad ds:= e^b\, dr ~. \ee

Finally, all fields are of a definite length dimension. The
dimensions of $r,\om$ are $1,-1$ respectively, while the
Lagrangian has dimension $-2$ as a curvature. The dimensions of the metric functions and fields are \\
\begin{center}
\begin{tabular}{|c|c|c|c|c|c|c|c|c|c|c|c|c|c|}
   \hline
  $e^{2a}$ & $e^{2b}$ & $e^{2c}$ &
  $h^S_{tt}$ & $h^S_{tr}$ & $h^S_{rr}$ & $h^S_{t}$ & $h^V_{t}$ & $h^S_{r}$ & $h^V_{r}$ & $h^S$ & $\tilde h^S$ & $h^V$ & $h^T$ \\\hline
  0 & 0 & 2 & 0 & 0 & 0 & 1 & 1 & 1 & 1 & 2 & 2 & 2 & 2 \\ \hline
\end{tabular}
\end{center}
One can easily check that our Lagrangian summarized in (\ref{TL}),
(\ref{VectorLagrangian}) and (\ref{ScalarLagrangian}) is of
dimension $-2$.
\subsection{Tensor sector}
 \label{TensorSector}

There is only one field in the tensor sector, namely $h^T$.
Naturally, it is gauge-invariant. The action simplifies by the
following field redefinition \be
 \phi^T:=e^{-2c}\, h^T ~,\ee
 which is nothing but the divergenceless part of the mixed
index components $h^i_j$. In terms of $\phi^T$ the tensor action is
given by \bea
 S^T &=& \int e^{a+(D-2)c}\, ds\, {\cal L}^T \non
 {\cal L}^T &=&-\half\, \bigl( \del_r\phi^T\, \del^r \phi^T +
 (C_2\, e^{-2c} + \om^2\, e^{-2a})\, \phi^{T^2}\bigr) ~,\label{TL}\eea
 where \be
 C_2:=l(l+D-3)
 \label{TensorLagrangian}\ee
 is a value of a quadratic Casimir operator in a scalar representation
with highest weight $l$ (for more details about various
representations see Appendix \ref{SphericalHarmonics}). We will also
find it useful to define \be\Ch:=C_2-(D-2)=(l+D-2)(l-1).\ee \emph{In
the expression (\ref{TensorLagrangian}) and throughout the rest of
the paper bilinear expressions such as $\phi^2$ must be understood
as $\phi\phi^*$ and the real part extracted}. (Even if the two
fields are different, the expression is the same for either
possibility of which field to conjugate).

\subsection{Vector sector} \label{VectorSector}

In the vector sector there are 3 fields: $h^V_t$, $h^V_r$ and $h^V$.
\footnote{In this subsection all fields will have a $V$ superscript.
It would have been possible to omit it, but we didn't do so in order
to avoid a possible confusion with scalar fields in the next
subsection with an otherwise identical notation. The same comment
applies to the $S$ superscript in the next subsection.} The vector
part of the action is \bea
 S^V &=& C_2\int e^{a+(D-2)c}\, ds\, {\cal L}^V \non
 {\cal L}^V &=& {\cal L}_2+{\cal L}_1+{\cal L}_0,\label{VectorLagrangian} \eea
 where ${\cal L}_2$ is a ``kinetic term", ${\cal L}_1$ is a term with one derivative w.r.t.
$r$, and ${\cal L}_0$ is a ``potential":  \be
 {\cal L}_2=-e^{-2a-2c}\,\del_r h_t^V\, \del^r h_t^V
 -\frac {\Ch}{4}e^{-4c}\, \del_r h^V\, \del^r h^V ~,\ee
 \be {\cal L}_1=-2i  \om\, e^{-2a-2c}\, \del^r h_t^V h^V_r+  \Ch
e^{-4c}\, \del^r h^V\, h^V_r~,\ee
\begin{multline}
~~~{\cal L}_0=-e^{-2c}\(\Ch
e^{-2c}+\om^2e^{-2a}\)\,h^{V}_rh^{V\,r}+4i\om\,e^{-a-3c}\,\sqrt{g^{rr}}\,h^V_rh^V_t\\-2\,\Ch
e^{a-5c}\sqrt{g^{rr}}\,h^V_rh^V-e^{-2a-4c}\(\Ch+2(D-3)e^{2a}\)\,h^{V^2}_t\\+i\Ch\om
e^{-2a-4c}\,h^V_th^V-\frac{\Ch}{4}e^{-2a-6c}\(e^{2c}\om^2-4e^{4a}+2(D-3)e^{2a}\)\,h^{V^2}\label{L0Vector}
\end{multline}

The gauge variations of the fields are (see (\ref{GaugeComponents})
for the definitions) \be
 \delta h^V_t=i\om\xi^V~,\qquad \delta h^V_r=\xi^{V\prime}-2c'\xi^V~,\qquad \delta h^V=2\xi^V ~.\ee
Altogether the number of fields and gauge functions is $n_F=3,\,
n_G=1$ and hence according to the general formula (\ref{n-dyn}) we
expect a single dynamic field ($3-2*1=1$) and one auxiliary field.

Since $h^V_r$ is an $r$-vector it is seen that its variation
includes $\xi^{V\prime}$ and hence it is ``derivatively-gauged'' in
the terminology of \cite{1dPert}. According to the general theorem
formulated and proved there and reviewed in section \ref{review} the
derivatively gauged fields decouple from the rest of the fields
after a certain shift in their definitions (completion of a square)
and the newly defined fields are auxiliary. An important role is
played by the quantity in the following definition \be
 \Dl^V := \Ch(l)\, e^{-2c(r)} + \om^2\, e^{-2a(r)} ~,\label{def-DV} \ee
 which comes from the auxiliary action, and in the scalar sector will be
seen to generalize to be essentially the determinant of the
auxiliary action (up to a multiplicative factor). It appeared in
(\ref{L0Vector}) as a coefficient of $h^{V^2}_r$. Now the shifted
field can be written as
\begin{multline}
H^V_r=- \Dl^V h^V_r + i\, \om\, e^{-a} (2 e^{-c} \sqrt{g_{rr}}
-e^{-a} \del_r)\, h_t^V
 -\half \Ch\, e^{a-2c} (2 e^{-c} \sqrt{g_{rr}} -e^{-a} \del_r)\, h^V
\end{multline}
and derivatives of this new field do not appear in the Lagrangian,
so this field is indeed auxiliary.

The remaining part of the Lagrangian contains only algebraically
gauged fields (fields whose variation does not contain
$\xi^{V\prime}$). Hence, in accordance with the general procedure of
\cite{1dPert} each gauge function can be used to eliminate one field
(``each gauge function shoots twice" -- the gauge function already
``produced'' an auxiliary field).

In our case we have single gauge function $\xi^V$, so we can
eliminate one algebraically gauged field out of two, or in other
words, we can write the Lagrangian in terms of a single gauge
invariant combination of fields. We define this dynamical field to
be \be
 \phi^V \equiv \phi_T^V = e^{-c}\,(\om\, h^V+2\,i\,h^V_t) ~. \label{phiV} \ee
  The Lagrangian then includes terms proportional to $\phi^{\prime2}$,
$\phi\phi'$, $\phi^2$ and $H_r^{V~2}$. The term $\phi \phi'$ can be
integrated by parts (it will change the coefficient of $\phi^2$).
The remaining Lagrangian decouples into a dynamic part $\cl_{Dyn}$
and an auxiliary (or algebraic part) $\cl_A$ \be
 \cl^V = \cl_{Dyn}^V + \cl_A^V \ee
where
 \bea
 \cl_{Dyn}^V &=& -\frac{C_2\, \Ch}{4}\, \om^2\, \[ \frac{e^{-2a}}{\Dl^V}\, \del_r \phi^V\, \del^r \phi^V
 +e^{-2a}\phi^{V^2} \] \label{FinalVectorLagrangian} \\
 \cl_A^V &=& - \frac{C_2\, e^{-2c}}{\Dl^V}\, H_r^V\, H^{V\,r} ~,\eea
where $\Delta^V$ is defined in (\ref{def-DV}).

\subsection*{Equivalence with Regge-Wheeler}

We shall now demonstrate that in the 4d case the EOM derived from
 the Lagrangian (\ref{FinalVectorLagrangian}) turns out to be
equivalent to the Regge-Wheeler equation \cite{RW} (see also
\cite{MartelPoisson}) after a certain canonical transformation which
essentially involves exchanging the field with its momentum. The
Cunningham-Price-Moncrief master field $\varphi_{CPM}$, in terms of
which the EOM is usually written, is connected to our field $\phi^V$
by \be
 \varphi_{CPM}=\frac{ie^{c}}{2\, C_2\, \Ch\,\Dl^V}\, \left. (\om H^V_r- C_2 \Ch\, \del_r \phi^V)\right|_{D=4} ~,\ee
 where the auxiliary $H^V_r$ can be put to $0$ due to its algebraic
EOM. Specializing (\ref{FinalVectorLagrangian}) to $D=4$ and
Schwarzschild coordinates (\ref{Schw-coord}) the dynamic part of our
action becomes \be
 S_{Dyn}^V = \int dr\, \frac{r^2}{2}\bigl(\frac{{\phi^{V^{\prime 2
}}}}{\Dl^V}+f^{-1}\phi^{V^2}\bigr) ~. \ee
 with $f=1-\frac{r_0}{r}$ (the unimportant overall factor was changed to simplify the expressions).

Before we perform the necessary canonical transformation we have to
turn first to the Hamiltonian. The conjugate momentum is $\pi^{V}
\equiv \delta S_{Dyn}/\delta(\del_r \phi^V)=\frac{r^2}{\Delta^V}\,
\phi^{V^{\prime}}$ and the Hamiltonian is \be
 {\cal H}=\frac12\bigl(\frac{\Delta^V}{r^2}\pi^{V^2}-r^2f^{-1}\phi^{V^2}\bigr) ~.\ee
 Now we perform a canonical transformation \be
 \tilde\phi^V=\frac{\pi^V}{r},\ \tilde {\pi}^V=-r\phi^V ~,\ee
 (in fact the new function $\tilde{\phi^V}$ is
equal to $\varphi_{CPM}$ up to a numerical factor). This
transformation is produced by the generating function
 $F(\phi^V,\tilde{\phi^V},r)=r\, \phi^V\, \tilde{\phi^V}$.
 In general, for a generating function of the form \be
  F=\eta(r)\phi\tl\phi \ee
  the canonical transformation is given by \be
 \tl\phi=\frac{\pi}{\eta},\qquad\tl\pi=-\eta\phi,\qquad\tl{\cal
 H}={\cal H}+\frac{\eta'}{\eta}\phi\tl\phi.\label{GeneralCanonicalTransformatin}\ee

Performing the canonical transformation gives $\tilde {\cal
H}=\frac12\bigl(\Dl^V\tilde{\phi}^{V^2}-f^{-1}\tilde
{\pi}^{V^2}\bigr)-\tilde{\phi}^V\tilde\pi^V/r$ and then Legendre
transforming back to a Lagrangian we find \be
 \tilde {\cal L}
 =-\frac f2\tilde{\phi}^{V^{\prime 2}}-\frac
fr\tilde{\phi}^V\tilde{\phi}^{V'}-
\frac{\Dl^V}{2}\tilde{\phi}^{V^2}-\frac{f}{2r^2}\tilde{\phi}^{V^2}.\ee
 The corresponding EOM is
\be\Box\tilde\phi^V=\bigl(\frac{ C_2 }{r^2}-\frac{3
r_0}{r^3}\bigr)\tilde\phi^V,\ee where $\Box=\partial_r(f\partial_r)
-f^{-1}\, \om^2$ is the $(r,t)$ Laplacian (or D'alembertian). This
is precisely the Regge-Wheeler equation.\footnote{Up to a change to
Lorentzian signature which amounts to $\om^2 \to -\om^2$.}

These considerations can be easily generalized to the case of
arbitrary dimension $D$ and any background gauge. Starting with the
Lagrangian (\ref{FinalVectorLagrangian}), we turn to the
Hamiltonian, perform a canonical transformation with the generating
function \be F=e^{\frac{D-2}{2}c}\phi^V\tl\phi^V\ee according to
(\ref{GeneralCanonicalTransformatin}) and transform back to the
Lagrangian. The result is
\begin{multline} \tl{\cal
L}^V_{Dyn}=e^{-(D-2)c}\partial_r\tphi^V\partial^{\,r}\tphi^V\\+
\frac14e^{-2a-Dc}\Bigl\{4e^{2c}\om^2+3(D-2)^2e^{4a}+\bigl(4C_2-2(D-1)(D-2)\bigr)e^{2a}\Bigr\}\tl\phi^2.\label{GeneralVectorLagrangian}
\end{multline}
In the Schwarzschild coordinates this final form of the action
encodes the master equation of
\cite{IshibashiKodama}\footnotemark[\value{footnote}] \be
 S^V = \int dr \( f\tl\phi^{V^{\prime2}} +
 \Bigl\{\frac{4C_2-2(D-1)(D-2)f}{4r^2f}-\frac{3(D-2)^2f}{4r^2}+\frac{\om^2}{f}\Bigr\}\tl\phi^{V^2} \)~.
 \label{SchwVectorLagrangian}
\ee where $f$ is defined in (\ref{Schw-coord}).

\subsection{Scalar sector}
 \label{ScalarSector}

In this section we apply the same procedure to the scalar sector. In
this sector there are 7 fields: $h^S_{tt}$, $h^S_{tr}$, $h^S_{rr}$,
$h^S_t$, $h^S_r, h^S$ and $\tilde h^S$. Their gauge variations are
\begin{gather}
\delta h^S_{tt}=2\(i\om\,\xi^S_t+a'e^{2a}\,\xi^{S\,r}\)~,\ \ \
\delta h^S_{tr}=\xi^{S\prime}_t-2a'\,\xi^S_t+i\om\,\xi^S_r~,\ \ \
\delta h^S_{rr}=2\(\xi^{S\prime}_r-b'\,\xi^S_r\)~,\non \delta
h^S_t=\xi^S_t+i\om\,\xi^S~,\qquad \delta
h^S_r=\xi^{S\prime}+\xi^S_r-2c'\,\xi^S~,\non \delta h^S=2c'e^{2c}\,
\xi^{S\,r}-\frac{2\,C_2 }{D-2}\,\xi^S~,\qquad \delta\tilde
h^S=2\xi^S\label{ScalarGauge}~.
\end{gather}
Altogether the number of fields and gauge functions is $n_F=7,\,
n_G=3$ and hence according to the general formula (\ref{n-dyn}) we
expect a single dynamic field ($7-2*7=1$) and 3 auxiliary fields.

Indeed, in the terminology of \cite{1dPert} the derivatively-gauged
fields are $h^S_{rr},\, h^S_{tr}$  and $h^S_r$, and the rest are
algebraically gauged, so the general theorem says that the
Lagrangian can be separated into an auxiliary part consisting of 3
fields $H^S_{rr}$, $H^S_{tr}$ and $H^S_r$ (they are 3
derivatively-gauged fields shifted by additional terms coming from a
completion to a square) and a dynamical part consisting of a single
gauge-invariant field $\phi^S$. We define \be
 \phi^S \equiv \phi_{tt}^S := h^S_{tt}-\frac{a'}{c'}e^{2(a-c)}h^S-\Bigl[\frac{ C_2
}{D-2}e^{2(a-c)}\frac{a'}{c'}+\om^2\Bigr]\tilde h^S-2\,i\,\om
h^S_t,\ee In terms of this field the Lagrangian can be written as
\bea
 S^S &=& \int e^{a+(D-2)c}\, ds\, {\cal L}^S \non
  \cl^S &=& \cl_{Dyn}^S + \cl_A^S \non
 {\cal L}_{Dyn}^{S} &=&\frac{1}{\Dl^S} \bigl(\hal\, \del_r \phi^S\, \del^r \phi^S
  +\hbt\, \sqrt{g^{rr}}\, \phi^S\phi^{S\prime}
 +\hgm\, \phi^{S^2}\bigr)
~, \label{ScalarLagrangian} \eea
 where ${\cal L}_A$ is an auxiliary Lagrangian
and
 (the expressions are organized by first expanding in powers of
$e^{2c}\om^2$, each coefficient expanded again in powers of $C_2$
and a final expansion in powers of $e^{2a}$).
\begin{multline}
\Dl^S=(D-2)^2\,e^{4c}\om^4+(D-2)(D-3)\Bigl[ C_2 (e^{2a}+1)-2(D-2)\,e^{2a}\Bigr]\,e^{2c}\om^2\\
+\frac{ C_2 }{4}(D-3)^2\Bigl[ C_2
(e^{2a}+1)^2-4(D-2)\,e^{2a}\Bigr]~,
\end{multline}

\begin{eqnarray*}
\hal &=&-\frac{C_2\Ch}{2}(D-2)(D-3) ~,\nonumber\\
\hbt &=& \frac
 {C_2}{4}e^{-3a-c}\Bigl\{2(D-2)\Bigl[2 C_2 -(D-2)\bigl((D-1)\,e^{2a}-D+3\bigr)\Bigr]\,e^{2c}\om^2\\
&&{}+2 C_2 ^{\,2}(D-3)(e^{2a}+1)- C_2
(D-2)(D-3)\,\Bigl[(3D-11)\,e^{4a}-2(D-6)\,e^{2a}-D+3\Bigr]\nonumber\\
&&{}+12(D-2)^2(D-3)^2e^{2a}(e^{2a}-1)\Bigr\}\nonumber,\\
\hgm &=& \frac
 {C_2}{8}e^{-4a-2c}\Biggl\{2(D-2)\Bigl[2 C_2
 -(D-2)\bigl((D-1)\,e^{2a}-D+3\bigr)\Bigr]\Bigl((D-5)\,e^{2a}-D+3\Bigr)\,
e^{2c}\om^2\\
&&{}+(D-3)\Bigl\{2 C_2 ^2\bigl((D-5)\,e^{4a}-D+3\bigr)\\
&&{}- C_2 (D-2)\Bigl[2(D-3)(D-5)\,e^{6a}-(3D^2-30D+67)\,e^{4a}-8(D-3)\,e^{2a}+(D-3)^2\Bigr]\\
&&{}+4(D-2)^2(D-3)^2\,e^{2a}(e^{2a}-1)^2\Bigr\}\Biggr\}\nonumber,
\end{eqnarray*}

The auxiliary Lagrangian is given by \be
 {\cal L}_A^S=\frac{1}{(\Dl^S)^2}\ H^{S\dagger}\, M\, H^S
\label{scalar-aux} \ee with
 \be
 H^S=\left(
        \begin{array}{c}
          H^S_{rr}\\
          H^S_{tr}\\
          H^S_r
        \end{array}
      \right)
      \ee (the auxiliary fields $H_{rr}$, $H_{tr}$ and $H_r$ are
      defined in the Appendix \ref{Auxiliary fields}) and $M$ is a
      $3\times 3$ Hermitian matrix given by
 \begin{multline} M=\\ \left(
 \begin{array}{ccc}
 \frac{(D-2)(D-3)}{2}\,e^{2a-4b} & i(D-2)\,e^{a-3b+c}\om &  -\frac {  (D-3)}{2}\,e^{a-3b-c}(e^{2a}+1)\\
  -i(D-2)\,e^{a-3b+c}\om & - C_2\, e^{-2b}  & i \om\, e^{-2b} \\
  -\frac {(D-3)}{2}\,e^{a-3b-c}(e^{2a}+1) & -i \om\, e^{-2b}  &\
  C_2^{-1}\, e^{-2b-2c}\bigl(2(D-3)\,e^{2a}-e^{2c}\om^2\bigr) \\
  \end{array}          \right) \label{defM}
\end{multline}

\subsection*{Equivalence with Zerilli}

We shall now show that just like in the vector case the standard
Zerilli equation is obtained after an analogous canonical
transformation. In $D=4$ and Schwarzschild coordinates the master
field of Zerilli (as given in \cite{MartelPoisson}) is \be
 \Psi_Z=\frac{r_0-r}{3r_0+r( C_2 -2)}\pi^S ~.\ee
  An action that yields Zerilli's equation is
\be S_Z=\int dr\,
\[ f\,\Psi_Z^{\,\prime^2}+V_Z(r,\om)\,\Psi_Z^{\,2} \]~,\ee
 where \be
V_Z(r,\om)=\frac{1}{\Lambda(r)^2}\Bigl[( C_2 -2)^2\,\bigl(\frac{ C_2
}{r^2}+\frac{3r_0}{r^3}\bigr) +\frac{9r_0^2}{r^4}\bigl( C_2
-2+\frac{r_0}{r}\bigr)\Bigr]+\frac{\om^2}{f}~,\ee
 and \be
 \Lambda(r)= C_2 -2+\frac{3 r_0}{r}~.\ee

Just like the case of the vector sector this Lagrangian is obtained
from ours by performing a canonical transformation with the
generating function \be F=-\frac{\Lambda(r)}{f}\phi^S\tilde{\phi^S}
\ee (and integrating by parts the term with one derivative).

Generalizing to arbitrary dimension and background gauge the
generating function is \be
 F=-e^{-2a+\frac{D-4}{2}c}\,\Lambda_D(r)~,\ee
 where \be
\Lambda_D(r)=\Ch+\frac{(D-1)(D-2)}{2}\,(1-e^{2a})~.\ee
 Then our usual procedure leads to the following action \be
 \tl S^S  = \int e^a\, ds \[ \del_r \tphi^S\, \del^r \phi^S + V(r)\,\tphi^{S^2} \]~,\ee
where \be V(r)=\frac{e^{-2c}}{\Lambda_D^2(r)}\, Q(r)
 +e^{-2a}\,\om^2~,\ee with
\begin{eqnarray*}
 Q(r) &=& \(\Ch^{\,3}+\al\, \Ch^{\,2}+\bt\, \Ch+\gm \) \\
 \al &=& \frac{D-2}{4}\Bigl\{3(D-6)e^{2a}-2(D-9)\Bigr\},\\
 \bt &=& \frac{(D-1)(D-2)}{4}(e^{2a}-1)\Bigl\{(2D^2-11D+18)e^{2a}+(D-1)(D-6)\Bigr\},\\
 \gm &=& -\frac{(D-1)^2(D-2)^2}{16}(e^{2a}-1)^2\Bigl\{(D-2)^2e^{2a}-2(D^2-7D+14)\Bigr\}.
 \label{albtgm}
\end{eqnarray*}
Using Schwarzschild coordinates, we obtain the following action \be
 \tl S^S = \int dr\, \[ f\tl\phi^{S^{\prime2}} + \Bigl\{\frac{Q(r)}{r^2\Lambda_D^2(r)}
 +f^{-1}\, \om^2 \Bigr\}\tl\phi^{S^2} \] ~,  \label{ScalarIK} \ee
which reproduces the equation obtained in
\cite{IshibashiKodama}.\footnotemark[\value{footnote}]
\section{Summary of results}
\label{summary}

In this section we collect our main results.

The background \Schw metric, allowing for an arbitrary choice of
the radial coordinate can be written as \be
 ds^2_{(0)}= -e^{2a(r)}\, dt^2 + e^{2b(r)}\, dr^2 + e^{2c(r)}\,
 d\Om^2_{D-2} ~, \label{Lor-Schw} \ee
 thereby defining the background fields $a, b$ and $c$.
 The perturbation field $h_{\mu\nu}$ is defined through \be
 g_{\mu\nu}=g_{\mu\nu}^{(0)} + h_{\mu\nu} ~.\ee

The usual Lorentzian \Schw metric (\ref{Lor-Schw}) has a Euclidean
version (\ref{background}) obtained simply by reversing the sign
of $g_{tt}$. Since the metric is static the transformation between
the two is quite straightforward and is effected by the analytic
continuation $t \to i\, t$. Whereas so far we used the Euclidean
signature (for the sake of a certain mathematical uniformity) in
this section we transform to the more physical Lorentzian
signature. The effect of the analytic continuation $t \to i\, t$
on our formulae is given by the following transformation \bea
 \om &\to& -i\, \om \non
 h_t &\to& -i\, h_t ~~;~ h_{tr} \to -i\, h_{tr} ~~;~ h_{tt} \to - h_{tt} \label{Euc-to-Lor} \eea

\presub {\bf The tensor sector.} The tensor sector includes a
single field and no gauge functions, so there is no issue of gauge
choice or decoupling.

The action in the tensor sector is given by \be
 S^T=\int e^{a+(D-2)c}\, ds\, \cl^T ~,\ee
 where \be
 ds:=e^b\, dr \ee
 is the 1d ``volume'' element and $e^{a+(D-2)c}\,
 ds\,=\int \sqrt{-g_{D}}\, d^Dx$ is the D-dimensional volume element
reduced over time and the sphere. The Lagrangian density is given
by \be
 \cl^T = \half \[ \om^2\, e^{-2 a}\, \phi^T \phi^{T*} - \del_r \phi^T\, \del^r \phi^{T*}
  - V^T(r)\, \phi^T \phi^{T*}  \] \ee
where \bi
 \item $\phi^T=\phi^T_{\om l}(r)$ is the traceless and divergenceless
 part of $h^i_j$;
 \item $~~V^T(r) =C_2(l)\, e^{-2c}~~$ $~~$ is the potential, and
 \item $C_2(l) = l(l+D-3)$.
\ei

\presub {\bf The vector sector.} In this sector we count 3 fields
and one gauge function, and therefore we have a single dynamic
field (the master field) and a single auxiliary field.

The action is given by \be
 S^V = C_2 \int e^{a+(D-2)c}\, ds\, ~\[ \cl^V_{Dyn} +\cl^V_A ~\] ~.\ee
 The auxiliary action is \be
 \cl^V_A= -\frac{C_2}{\Delta^V}\, H_r^V\, H^{Vr*} \ee
 where
 \be
 \Delta^V =  \Ch(l) e^{-2c} - \om^2\, e^{-2a} \ee
and \be
 \Ch := (l-1)(l+D-2) ~.\ee
 The auxiliary field is an $r$-vector that can be expressed in terms of the
 fields $h_{\mu\nu}$ as follows \be
 H_r^V = -\Dl^V\, h_r^V - i\, \om\, e^{-a} (2 e^{-c} \sqrt{g_{rr}} -e^{-a} \del_r)\, h_t^V
 -\half \Ch\, e^{a-2c} (2 e^{-c} \sqrt{g_{rr}} -e^{-a} \del_r)\, h^V ~.\ee

For the dynamic part we have two expressions. The first is the one
that naturally emerges from the auxiliary/dynamic method and it is
given by \be
 \cl^V_{Dyn} = -\frac {C_2\, \Ch}{4}\, \om^2\, \[ \frac{e^{-2a}}{\Dl^V}\, \del_r \phi_t^V\, \del^r\phi_t^{V*}
 + \phi_t^V\, \phi^{Vt*}\] \ee
 where the dynamic field is \be
 \phi^V_t=e^{-c}\, \( \om h^V+2 i\, h_t^V \) \ee
 (after dropping an overall ($-i$) factor that comes from the
 analytic continuation of (\ref{phiV})).

The second expression encodes the standard Regge-Wheeler master
equation (as generalized by Kodama-Ishibashi). It is \be
 S^V = \int e^a\, ds\, \[ \om^2\, e^{-2a}\, \tphi^V \tphi^{V*}
  - \del_r \tphi^V\, \del^r \tphi^{V*} -V^V(r)\, \tphi^V
  \tphi^{V*} \] ~,\ee
  where the potential in the vector sector is \be
  V^V(r) =\( \frac{3}{4}(D-2)^2\, e^{2a}+ C_2 - \frac{(D-1)(D-2)}{2}
  \)\, e^{-2c} ~,\ee
  and the new dynamic field, which is essential the one
  canonically conjugate to $\phi_t^V$ is \be
 \tphi^V=\frac{\exp(\frac{D-2}{2}c -b -a)}{\Dl^V}\, \del_r\phi_t^V
 ~. \ee

\presub {\bf The scalar sector.} In this sector we find 7 fields
and 3 gauge functions, and therefore we still have a single
dynamic field (the master field) and this time there are 3
auxiliary fields.

The action is given by \be
 S^V = C_2 \int e^{a+(D-2)c}\, ds\, ~\[ \cl^S_{Dyn} +\cl^S_A ~\] ~.\ee
The auxiliary sector includes the $r$-tensor $H_{rr}^S$ as well as
the two $r$-vectors $H_{rt}^S$ and $H_{r}^S$. Their action is a
straightforward analytic continuation of
(\ref{scalar-aux},\ref{defM}) according to the rules
(\ref{Euc-to-Lor}).

The first expression for the dynamic action $\cl^S_{Dyn}$ in terms
of the master field $\phi^S_{tt}$ is quite involved and can be
gotten by an analytic continuation of (\ref{ScalarLagrangian}).

The second expression produces the standard Zerilli master
equation (as generalized by Kodama-Ishibashi). It is \be
 S^S = \int e^a\, ds\, \[ \om^2\, e^{-2a}\, \tphi^S \tphi^{S*}
  - \del_r \tphi^S\, \del^r \tphi^{S*} -V^S(r)\, \tphi^S
  \tphi^{S*} \] ~,\ee
  where the potential in the scalar sector is \be
  V^S(r) = \frac{e^{-2c}}{\Lambda_D^2} \(\Ch^3+ \al\, \Ch^2+ \bt\, \Ch + \gm \)
  \ee together with \be
  \Lambda_D = \Ch(l) + \frac{(D-1)(D-2)}{2}(1-e^{2a}) ~,\ee
 and the coefficients $\al, \bt, \gm$ which depend only on $D$ and $e^{a(r)}$ are given
in (\ref{albtgm}). The new dynamic field $\tphi^S$ is essentially
the momentum canonically conjugate to $\phi_{tt}^S$.

\subsubsection*{Acknowledgements}

We thank Michael Smolkin for discussions. This work was finalized
during the workshop ``Einstein's Gravity in Higher Dimensions",
Jerusalem, 18–22 February 2007. This research is supported in part
by The Israel Science Foundation grant no 607/05, DIP grant H.52,
EU grant MRTN-CT-2004-512194 the Einstein Center at the Hebrew
University, and by The Binational Science Foundation BSF-2004117.

\newpage
\appendix
\section{Derivation of the one-dimensional action}
\label{1DAction}

The action for gravitational waves around a general background was
given in (\ref{MultiDAction}). In this section we reduce it to
obtain a 1D action. First of all we need to separate the indices
into sets according to their behavior w.r.t. rotations: $t$ and $r$
are denoted by $\al,\, \bt,\dots$ while the angular coordinates are
denoted by Latin letters $i,\, j,\dots$. In addition, we have to
take into account the fact that covariant derivatives on the
$(D-2)$-dimensional sphere and in the ambient space differ in
Christoffel symbols with at least one index in $(r,t)$. In fact the
only non-vanishing Christoffel symbols are of the form
$\Gamma^{\gamma}_{\alpha\beta}$, $\Gamma^j_{i\al}$, $\Gm^{\al}_{ij}$
and $\Gm^i_{jk}$. We denote the covariant derivatives on a sphere by
$\Nt_i$. With indices separated in this way and after introducing
covariant derivatives on a sphere the action gets the form \be
S=\int
d^Dx\sqrt{g}\Bigl(A_3+A_2+A_1+A_0+\frac{1}{2(D-2)}(B_3+B_2+B_1+B_0)\Bigr),\ee
where the $A$-terms and $B$-terms come from the first and second
terms in (\ref{MultiDAction}),
 and subscripts highlight the number of $(t,r)$ indices in the
``parent" term. We have (where in what follows a comma means a usual
partial derivative w.r.t. $r$, rather than a covariant one)
\begin{gather}
A_3=\hb_{\bt\gm ,\al}\hb_{\bt\gm ,\al}-2\,\hb_{\bt\gm
,\al}\hb_{\al\gm ,\bt}+ 4\,\Gm^{\mu}_{\al\bt}\hb_{\mu\gm}\hb_{\al\bt
,\gm}-
4\Gm^{\mu}_{\al\bt}\Gm^{\nu}_{\bt\gm}\hb_{\mu\gm}\hb_{\nu\al},\non\non
A_2=\Nt_i\hb_{\al\bt}\Nt_i\hb_{\al\bt}+2\,\hb_{i\bt,\al}\hb_{i\bt,\al}-2\,\hb_{i\bt,\al}
\hb_{i\al,\bt}-4\,\Nt_i\hb_{\al\bt}\hb_{i\bt,\al}\non
+8\,\Gm^j_{i\al}\hb_{j\bt}\hb_{i\al,\bt}+4\,\Gm^{\mu}_{\al\bt}\hb_{i\mu}\Nt_i
\hb_{\al\bt}-4\,\Gm^j_{i\al}\Gm^k_{i\bt}\hb_{j\bt}\hb_{k\al}
-8\,\Gm^j_{i\bt}\Gm^{\mu}_{\al\bt}\hb_{j\al}\hb_{i\mu},\non\non
A_1=2\,\Nt_i\hb_{j\al}\Nt_i\hb_{j\al}-2\,\Nt_i\hb_{j\al}\Nt_j\hb_{i\al}-4\,\hb_{ij,\al}\Nt_i
\hb_{j\al}+\hb_{ij,\al}\hb_{ij,\al}\non
+4\,\Gm^{\bt}_{ij}\hb_{\al\bt}\hb_{ij,\al}+8\,\Gm^k_{i\al}\hb_{kj}\Nt_j\hb_{i\al}-
8\,\Gm^{\al}_{ij}\Gm^k_{j\bt}\hb_{ki}\hb_{\al\bt}-4\,\Gm^k_{i\al}\Gm^{l}_{j\al}\hb_{kj}\hb_{li},\non\non
A_0=\Nt_i\hb_{jk}\Nt_i\hb_{jk}-2\,\Nt_i\hb_{jk}\Nt_j\hb_{ik}+4\,\Gm^{\al}_{ij}\hb_{k\al}
\Nt_k\hb_{ij}-4\,\Gm^{\al}_{ij}\Gm^{\bt}_{jk}\hb_{k\al}\hb_{i\bt},\non\non
B_3=\hb_{\bt\bt,\al}\hb_{\gm\gm,\al},\non
B_2=2\,\hb_{\bt\bt,\al}\hb_{ii,\al}+\Nt_i\hb_{\al\al}\Nt_i\hb_{\bt\bt},\non
B_1=\hb_{ii,\al}\hb_{jj,\al}+2\,\Nt_i\hb_{\al\al}\Nt_i\hb_{jj}\non
B_0=\Nt_i\hb_{jj}\Nt_i\hb_{kk}\nonumber.
\end{gather}
In these expressions the summation over the repeated indices is
assumed to be done with the corresponding components of the inverse
metric, which is diagonal; they are not written explicitly in order
not to clutter the notation. So, for example, $\hb_{\al\al}$ should
be understood as $g^{(0)\al\bt}\hb_{\al\bt}$, $\hb_{ii}$ as
$g^{(0)ij}\hb_{ij}$ etc. Note that this action is valid for
quadratic perturbations around any background of the form $\IS^m
\times X$, where $X$ and $m$ are arbitrary.

The explicit expressions for the inverse metric and the Christoffel
symbols in the background (\ref{background}) are
\begin{gather}
g^{(0)tt}=e^{-2a},\ g^{(0)rr}=e^{-2b},\ g^{(0)ij}=e^{-2c}\tl{g}^{ij},\\
\Gm^t_{tr}=a',\ \Gm^r_{tt}=-a'e^{2(a-b)},\ \Gm^r_{rr}=b',\
\Gm^j_{ri}=c'\delta^j_i,\ \Gm^r_{ij}=-c'e^{2(c-b)}\tl{g}_{ij},
\end{gather}
where $\tl{g}_{ij}$ is a metric on a unit $D-2$-dimensional sphere.

Next we separate the $\al,\bt$ indices according to
$\al\rightarrow(t,r)$, express $\hbar$'s in terms of $h$'s according
to the definition (\ref{hbar}), expand the fields as in
(\ref{Expansions}) and calculate angular integrals with the formulae
of Appendix (\ref{SphericalHarmonics}). In this way one arrives at
the action described in subsections
\ref{TensorSector}-\ref{ScalarSector}.

\section{Auxiliary fields in the scalar sector}
\label{Auxiliary fields}

In this appendix we give the precise definitions of the auxiliary
fields in the scalar sector. As mentioned in section
\ref{ScalarSector} there are 3 such fields that come from a
completion to a square of fields $h^S_{rr}$, $h^S_{tr}$ and $h^S_r$.
We called these fields $H^S_{rr}$, $H^S_{tr}$ and $H^S_r$. They are
invariant w.r.t. the gauge transformations (\ref{ScalarGauge}) and
appear in the Lagrangian of the scalar sector in an algebraic way
(without derivatives).

The first of these fields is given by  \be
H^S_{rr}:=e^{-3a-c+2b}\(\al_1h^S_{rr}+\al_2h^S_{tt}+\al_3h^S_t+\al_4h^S+\al_5\tilde
h^S+\al_6h^{S\prime}_{tt}+\al_7h^{S\prime}_t+\al_8h^{S\prime}+\al_9\tilde
h^{S\prime}\),\ee where the expressions that follow for the
coefficients are organized (as in section \ref{ScalarSector}) by
first expanding in powers of $e^{2c}\om^2$, each coefficient
expanded again in powers of $C_2$ and a final expansion in powers of
$e^{2a}$.
 $$ \al_1 = e^{2(a-b)}\, \Dl^S \qquad \qquad $$
\begin{subequations} \begin{multline*}
\al_2=\frac{i\al_3}{2\om}=-\frac{ C_2}{2}
(D-2)\Bigl[(D-5)e^{2a}-D+3\Bigr]e^{2c}\om^2\\- \frac{C_2}{4}
(D-3)(e^{2a}-1)\Biggl\{ C_2 \Bigl[(D-5)e^{2a}+D-3\Bigr]
-4(D-2)(D-3)e^{2a}\Biggr\};\\
\end{multline*}\end{subequations}
\begin{subequations} \begin{multline*}
\al_4=\frac 14 e^{-2c}\Biggl\{-2(D-2)^2\Bigl[(D-5)e^{2a}-D+3\Bigr]e^{4c}\om^4-(D-2)(D-3)e^{2c}\om^2\times\\
\Bigl\{\Bigl[ C_2 (D-13)-4(D-2)(D-5)\Bigr]e^{4a}+2\Bigl[ C_2 +2(D-2)(D-3)\Bigr]e^{2a}- C_2 (D-3)\Bigr\}\\
+4 C_2 (D-3)^2\Ch e^{6a}\Biggr\};\\
\end{multline*}\end{subequations}
\begin{subequations} \begin{multline*}
\al_5=C_2\Ch \frac{D-3}{D-2} e^{4a-2c}\Bigl\{2(D-2)e^{2c}\om^2+
C_2 (D-3)e^{2a}\Bigr\};\\
\end{multline*}\end{subequations}
\begin{subequations} \begin{multline*}
\al_6=\frac{i\al_7}{2\om}=-\frac{C_2}{2}e^{a-b+c}\Biggl\{2(D-2)e^{2c}\om^2+(D-3)\Bigl[
C_2 (e^{2a}+1)-2(D-2)e^{2a}\Bigr]\Biggr\};\\
\end{multline*}\end{subequations}
\begin{subequations} \begin{multline*}
\al_8=-\frac 12 e^{a-b-c}\Biggl\{2(D-2)^2e^{4c}\om^4+(D-2)(D-3)\Bigl[ C_2 (3e^{2a}+1)-4(D-2)e^{2a}\Bigr]e^{2c}\om^2\\
+ C_2 \Ch(D-3)^2(e^{4a}+e^{2a})\Biggr\};\\
\end{multline*}\end{subequations}
\begin{subequations} \begin{multline*}
\al_9=-\frac{C_2\Ch}{2} \frac{D-3}{D-2}
e^{3a-b-c}\Bigl\{2(D-2)e^{2c}\om^2+ C_2
(D-3)(e^{2a}+1)\Bigr\}.\\
\end{multline*}\end{subequations}

The second auxiliary field is given by \be
 H^S_{tr}:= e^{-a-c}\,\(\bt_1\, h^S_{tr} +\bt_2\, h^S_{tt} +\bt_3\, h^S_t +\bt_4\, h^S+\bt_5\, \tilde
 h^S +\bt_6\, h^{S\prime}_{tt} +\bt_7\, h^{S\prime}_t +\bt_8\, h^{S\prime}+\bt_9\, \tilde h^{S\prime}\), \ee
where the coefficients are
 $$ \bt_1=\Dl^S $$
\begin{subequations} \begin{multline*}
\bt_2=\frac i4
e^{-a+b+c}\om\Biggl\{2(D-2)(e^{2a}-1)\Bigl[(D-2)(D-3)- C_2
\Bigr]e^{2c}\om^2\\
-(D-3) \Bigl\{ C_2 ^2(e^{2a}+1)- C_2
(D-2)\Bigl[2(D-4)e^{4a}-(D-7)e^{2a}-D+3\Bigr]\\
+4(D-2)^2(D-3)(e^{4a}-e^{2a})\Bigr\}\Biggr\};\\
\end{multline*}\end{subequations}
\begin{subequations} \begin{multline*}
\bt_3=-\frac{C_2}{4}e^{-a+b-c}\Biggl\{4(D-2)e^{4c}\om^4\\
+2(D-3)\Bigl[ C_2 (e^{2a}-1)+(D-2)(2e^{4a}+(D-7)e^{2a}-D+3)\Bigr]e^{2c}\om^2\\
+(D-3)^3(e^{2a}-1)\Bigl[ C_2
(e^{2a}+1)^2-4(D-2)e^{2a}\Bigr]\Biggr\};\\
\end{multline*}\end{subequations}
\begin{subequations} \begin{multline*}
\bt_4=-\frac i4 e^{-a+b-c}\om\Biggl\{2(D-2)^2e^{4c}\om^4\\
+(D-2)(D-3)\Bigl\{ C_2 (3e^{2a}+1)+(D-2)\Bigl[2e^{4a}+(D-9)e^{2a}-D+3\Bigr]\Bigr\}e^{2c}\om^2\\
+\frac{(D-3)^2}{2}\Bigl\{2 C_2 ^2(e^{4a}-e^{2a})\\
+C_2 \Bigl[(D+5)(D-2)e^{6a}+(D-2)(D-13)e^{4a}-(D-1)(D-2)e^{2a}-(D-2)(D-3)\Bigr]\\
-4(D-2)^2e^{2a}(2e^{4a}+(D-5)e^{2a}-D+3)\Bigr\}\Biggr\};\\
\end{multline*}\end{subequations}
\begin{subequations} \begin{multline*}
\bt_5=-\frac {i\om}{4}
\Ch(D-3)e^{a+b-c}\Biggl\{2e^{2c}\om^2+\frac{D-3}{D-2}\Bigl[ C_2
(e^{2a}+1)+4(D-2)(e^{4a}-e^{2a})\Bigr]\Biggr\};\\
\end{multline*}\end{subequations}
\begin{subequations} \begin{multline*}
\bt_6=\frac{i(D-2)}{4\om}\Biggl\{2(D-2)e^{4c}\om^4+(D-3)\Bigl[ C_2
(3e^{2a}+1)-4(D-2)e^{2a}\Bigr]e^{2c}\om^2\Biggl\};\\
\end{multline*}\end{subequations}
\begin{subequations} \begin{multline*}
\bt_7= \frac{C_2}{4}
(D-3)\Biggl\{2(D-2)(e^{2a}-1)e^{2c}\om^2-(D-3)\Bigl[ C_2
(e^{2a}-1)^2-4(D-2)e^{2a}\Bigr]\Biggr\};\\
\end{multline*}\end{subequations}
\begin{subequations} \begin{multline*}
\bt_8=\frac i4
\om(D-2)(D-3)(e^{2a}-1)\Biggl\{(D-2)e^{2c}\om^2+\frac{D-3}{2}\Bigl[
C_2 (3e^{2a}+1)-4(D-2)e^{2a}\Bigr]\Biggr\};\\
\end{multline*}\end{subequations}
\begin{subequations} \begin{multline*}
\bt_9=\frac{iC_2}{2}\Ch(D-3)^2e^{4a}\om.\\
\end{multline*}\end{subequations}

The third auxiliary field is  \be
 H^S_{r}=C_2\,e^{-a-c}\(\gm_1h^S_r+\gm_2h^S_{tt}+\gm_3h^S_t+\gm_4h^S+\gm_5\tilde
h^S+\gm_6h^{S\prime}_{tt}+\gm_7h^{S\prime}_t+\gm_8h^{S\prime}+\gm_9\tilde
h^{S\prime}\),\ee where the coefficients are
 $$ \gm_1=\Dl^S $$
\begin{subequations} \begin{multline*}
\gm_2=\frac{i\gm_3}{2\om}=-\frac 14 e^{-a+b+c}\Biggl\{2(D-2)\Bigl[
C_2 -2(D-2)e^{2a}\Bigr]e^{2c}\om^2\\+ C_2 (D-3)\Bigl\{ C_2
(e^{2a}+1) -(D-2)e^{2a}\Bigl[(D-3)e^{2a}-D+5\Bigr]\Bigr\}\Biggr\};\\
\end{multline*}\end{subequations}
\begin{subequations} \begin{multline*}
\gm_4=-\frac 14
e^{-a+b-c}\Biggl\{2(D-2)^2e^{4c}\om^4\\+(D-2)(D-3)\Bigl\{ C_2
(3e^{2a}+1)
-(D-2)e^{2a}\Bigl[(D+1)e^{2a}-D+3\Bigr]\Bigr\}e^{2c}\om^2\\
+ C_2 (D-3)^2\Bigl[ C_2
(e^{2a}+1)+(D-2)(e^{4a}-3e^{2a})\Bigr]\Biggr\};\\
\end{multline*}\end{subequations}
\begin{subequations} \begin{multline*}
\gm_5=-\frac{\Ch}{4}(D-3)\,e^{a+b-c}\Biggl\{2\Bigl[ C_2 -4(D-2)e^{2a}\Bigr]e^{2c}\om^2\\
+ C_2 \frac{D-3}{D-2}\Bigl[ C_2
(e^{2a}+1)-4(D-2)e^{2a}\Bigr]\Biggr\};\\
\end{multline*}\end{subequations}
\begin{subequations} \begin{multline*}
\gm_6=\frac{i\gm_7}{2\om}=\frac{2e^{2c}\,\gm_8}{(D-3)\,(e^{2a}-1)}=-\frac{D-2}{4}e^{2c}\Bigl\{2(D-2)e^{2c}\om^2-
C_2 (D-3)(e^{2a}-1)\Bigr\};\\
\end{multline*}\end{subequations}
\begin{subequations} \begin{multline*}
\gm_9=-\frac{\Ch}{2}(D-3)\,e^{2a}\Bigl\{2(D-2)e^{2c}\om^2+ C_2
(D-3)\Bigr\};\\
\end{multline*}\end{subequations}

\section{Multidimensional spherical harmonics}
\label{SphericalHarmonics}

\subsection*{General theory}

We consider tensor spherical harmonics of various ranks on a $d-1$
dimensional sphere, where in our application $d=D-1$. We give here
the properties that we need, while furthermore an explicit
construction of the harmonics can be found, for example, in
\cite{Higuchi:1986wu}. Spherical harmonics transform in an
irreducible representation of $SO(d)$. These representations are
labeled by Dynkin indices $[l_1...l_n]$ where $n=[d/2]$ is the rank
of the group $SO(d)$ and each function within the representation is
further labeled by a ``weight vector'' $[m_1...m_n]$. Henceforth we
shall denote the multi-index $m_1...m_n$ collectively as $m$.

In this context it is necessary to distinguish two types of tensor
types. The first tensor-type is with respect to local
diffeomorpshims under $GL(d)$ and it will be denoted as a
$\Om$-scalar-field or $\Om$-vector-field etc. The second tensor type
is with respect to the $SO(d)$ global isometries: we will refer to
$[l,0,...0]$ as an $SO(d)$-scalar, to $[l,1,...0]$ as an
$SO(d)$-vector and to $[l,2,...0]$ as an $SO(d)$-tensor.

First, there are ($\Om$) scalar spherical harmonics $Y^S_{~l m}$
that transform under rotations of a sphere in the $SO(d)$-scalar
representation $[l,0,...0]$. These functions are all eigenfunctions
of the Laplace-Beltrami operator $\Box\equiv\N_i\N^i$ \be
 \Box Y^S_{~ l m}=-C_2\, Y^S_{~l m}~,\label{ScalarBox}\ee
 where $C_2:=l(l+d-2)$ is the eigenvalue of the quadratic Casimir operator
in the $SO(d)$-scalar representation $[l,0,...0]$. It is a
generalization of the well known formula for $d=3$ spherical
harmonics $\Box Y=-l(l+1)Y$.

Then there are ($\Om$) vector spherical harmonics, which belong to
two families according to the Hodge decomposition. The first family
consists of $\Om$-vector-fields of the following gradient form \be
 Y^S_{i,~ l m} = \N_i Y^S_{~l m} ~.\ee
These are $SO(d)$-scalars which explains the superscript $S$ (they
belong to the same type of $SO(d)$ representation $[l,0,...0]$ as
the $\Om$-scalar-fields). In view of (\ref{ScalarBox}) they obey \be
 \N_i Y^{S~i}_{~ l m}=-C_2\, Y^S_{~l m}~,\label{VectorTrace} \ee
 The functions in the second family of ($\Om$) vector harmonics
$Y^V_{i,~ l m}$ belong to the $SO(d)$-vector representation
$[l,1,...0]$ and are divergenseless: \be\N_i Y^{V~i}_{~ l m}=0~.\ee
The action of the Laplace-Beltrami operator on $Y^S_i$ can be
calculated from the definition: \be
 \Box Y^S_{i,~ l m} = -(C_2 -d+2)\, Y^S_{i,~ l m} ~.
 \ee
 whereas for $Y^V_i$ the general theory presented in
\cite{Higuchi:1986wu} gives \be
 \Box Y^V_{i,~ l m} = -(C_2 -1)\, Y^V_{i,~ l m} ~.
 \ee

Then there are ($\Om$) symmetric tensor spherical harmonics, which
come in 4 families. First of all, as $g_{ij}$ and $\N_i$ are
invariant tensors, it is obvious that the $\Om$-tensor valued
functions $g_{ij}\, Y^S_{~l m}$ and $\N_i\N_jY^S_{~l m}$ transform
in the $SO(d)$ -scalar representation $[l,0,...0]$. However they are
not orthogonal w.r.t. to the inner product
 \be <Y_1|Y_2> \equiv \int\limits_{S^{d-1}}Y_{1,ij}Y^{ij}_2d\Om\label{InnerProduct}~.\ee
  In order to orthogonalize this system of functions we define two
($\Om$) tensor spherical harmonics by \be
 Y^S_{ij,~ l m} := g_{ij}\, Y^S_{~l m}~,\qquad \tl
 Y^S_{ij,~ l m} := \bigl(\N_i\N_j+\frac{ C_2 }{d-1}\bigr)Y^S_{~l
 m}~.\label{OrthogonalHarmonics}\ee
Then there are spherical harmonics of the gradient form \be
 Y^V_{ij,~ l,m} \equiv \frac12(\N_i Y^V_{j,~ l m}+\N_j Y^V_{i,~
l m}) ~, \ee
 which belong to the vector representation $[l,1,0...0]$
and are traceless because of (\ref{VectorTrace}). Finally there is
the family of ($\Om$) tensor spherical harmonics $Y^T_{ij,~l m}$
which belong to the symmetric traceless tensor representation
$[l,2,0...0]$ and is divergenceless. The action of the
Laplace-Beltrami operator on $Y^S$, $\tl Y^S$ and $Y^V$ can be
calculated directly from their definitions:
\begin{gather}
 \Box Y^S_{ij,~ l m}=-C_2Y^S_{ij,~ l m}~,\non
 \Box \tl Y^S_{ij,~ l m}=-(C_2-2d+2)\tl Y^S_{ij,~ l m}~,\non
 \Box Y^V_{ij,~ l m}=-(C_2-d-1)Y^V_{ij,~ l m}\label{Boxes}~,
\end{gather}
 and the action on
$Y^T$ is given by (see \cite{Higuchi:1986wu})\be \Box Y^T_{ij,~ l
m}=-(C_2 -2)\, Y^T_{ij,~ l m} ~. \label{TensorBox}\ee Divergences of
tensor harmonics are given by
\begin{gather}
\N_i Y^{S\,ij}=Y^{S\,j}~,\qquad \N_i\tl
Y^{S\,ij}=-\frac{d-2}{d-1}(C_2-(d-1))\, Y^{S\,j}~,\nonumber\\
\N_iY^{V\,ij}=-\frac{C_2-(d-1)}{2}\, Y^{V\,j}~, \qquad
\N_iY^{T\,ij}=0~\label{Divergences}.
\end{gather}

\subsection*{Orthogonality and normalization}

The spherical harmonics constructed above are orthogonal to each
other. For harmonics belonging to different irreps it follows from
the usual theorems about unitary representations, and $Y^S_{ij}$ and
$\tl Y^S_{ij}$ are orthogonal by construction. Now we calculate the
norms of various spherical harmonics, where in fact the first line
in each group is just the definition of the appropriate
normalization.

\begin{itemize}
\item{Scalar}\be \int d\Om Y^S_{lm}Y^{S*}_{l'm'}=\delta_{ll'}\delta_{mm'}~.\label{ScalarNorm}\ee
\item{Vector} \bea
 \int d\Om\, Y_{i\:lm}^VY_{l'm'}^{V*\:i} &=& C_2\, \delta_{ll'}\delta_{mm'}~, \non
 \int d\Om\, Y_{i\:lm}^SY_{l'm'}^{S*\:i} &=& C_2\,
 \delta_{ll'}\delta_{mm'} ~. \eea
\item{Tensor}
\begin{gather}
\int d\Om\, Y_{ij\:lm}^SY_{l'm'}^{S*\:ij}=(d-1)\,
\delta_{ll'}\delta_{mm'}~,\non \int d\Om\, \tl Y_{ij\:lm}^S\tl
Y_{l'm'}^{S*\:ij}=\frac{d-2}{d-1}\, C_2 ( C_2 -d+1)\,
\delta_{ll'}\delta_{mm'}~,\non \int d\Om\,
Y_{ij\:lm}^VY_{l'm'}^{V*\:ij}=\frac { C_2 }{2}( C_2 -d+1)\,
\delta_{ll'}\delta_{mm'}~,\non \int d\Om\,
Y_{ij\:lm}^TY_{l'm'}^{T*\:ij}=\delta_{ll'}\delta_{mm'}~\label{TensorNorm}.
\end{gather}
\end{itemize}

\subsection*{Spherical harmonics on ${\bf S}^2$}
The case of spherical harmonics on ${\bf S}^2$ is special. The rank
of $SO(3)$ is 1, so ``weight vectors" consist of a single number,
and therefore all harmonics, no matter if they are scalar, vector
etc, belong to the same representations of $SO(3)$. Scalar spherical
harmonics are labelled by single Dynkin index $l$ and a single
weight $m$: $Y_{lm}$. Vector spherical harmonics can be built out of
scalars with the help of invariant tensors. There are 2 independent
invariant tensors: the covariant derivative on the sphere $\N_i$ and
the Levi-Civita tensor $\epsilon_{ij}$. One can build 2 families of
vector harmonics: \be Y^{(+)}_{i\,lm}=\N_iY_{lm'}~,\qquad
Y^{(-)}_{i\,lm}=\epsilon_i^{\:j}\N_jY_{lm'}~\label{2DVectorHarmonics}.\ee
The sign in the superscript denotes the spatial parity of the
harmonic with $l=0$. Higher-dimensional spherical harmonics $Y^S_i$
and $Y^V_i$ are generalizations of $Y^{(+)}$ and $Y^{(-)}$
correspondingly.

One can continue the process of covariant differentiation and
multiplication by $\epsilon_{ij}$ to build tensor spherical
harmonics of higher ranks. For example, one gets four families of
rank 2 tensor spherical harmonics:
\begin{gather}
Y^{(1)}_{lm\:ij}=\N_i\N_jY_{lm}~,\qquad Y^{(2)}_{lm\:ij}=\epsilon_i^{\ k}\N_k\N_jY_{lm}~,\nonumber\\
Y^{(3)}_{lm\:ij}=\epsilon_j^{\ k}\N_i\N_kY_{lm}~,\qquad
Y^{(4)}_{lm\:ij}=\epsilon_i^{\ k}\epsilon_j^{\ n}\N_k\N_nY_{lm}~,
\end{gather}
8 families at rank 3 etc. For our purposes we need vector and
symmetric rank 2 tensor harmonics. Out of four rank 2 tensor
harmonics one can build three symmetric ones: \be
Y^{(+1)}_{lm\:ij}=\N_i\N_jY_{lm}~,\qquad
Y^{(+2)}_{lm\:ij}=\epsilon_{(i}^{\ k}\epsilon_{j)}^{\
n}\N_k\N_nY_{lm}~,\qquad Y^{(-)}_{lm\:ij}=\epsilon_{(i}^{\
k}\N_{j)}\N_kY_{lm}~,\label{NonorthogonalTensorHarmonics}\ee
 where the first two have even parity while the last one is odd.

The spherical harmonics $Y^{(+1)}_{lm\:ij}$ and $Y^{(+2)}_{lm\:ij}$
in (\ref{NonorthogonalTensorHarmonics}) are not orthogonal w.r.t.
the inner product (\ref{InnerProduct}) (harmonics of the opposite
parity are obviously orthogonal). To orthogonalize them we notice
first that \be Y_{ij}^{(+1)}+Y_{ij}^{(+2)}\equiv -g_{ij}l(l+1)Y~.\ee
Therefore the orthogonal linear combinations are similar to
(\ref{OrthogonalHarmonics}):
\begin{gather}
Y_{lm\:ij}^{(+)}=g_{ij}Y_{lm}~,\nonumber\\ \tl
Y^{(+)}_{lm\:ij}=\N_i\N_jY_{lm}-\frac12g_{ij} \Box
Y_{lm}\equiv\N_i\N_jY_{lm}+\frac{l(l+1)}{2}g_{ij}Y_{lm}~.
\end{gather}
In higher dimensions the harmonics $Y^{(+)}_{ij}$, $\tl
Y^{(+)}_{ij}$ and $Y^{(-)}_{ij}$ become $Y^S_{ij}$, $\tl Y^S_{ij}$
and $Y^V_{ij}$. There is no analog of $Y^T_{ij}$ on ${\bf S}^2$.

The actions of Laplace-Beltrami operators on various spherical
harmonics are directly obtained from the higher-dimensional formulas
(\ref{Boxes}), their divergences from (\ref{Divergences}) and the
normalizations from (\ref{ScalarNorm} - \ref{TensorNorm}).


\newpage
\begin{thebibliography}{99}

\bibitem{1dPert}
  B.~Kol,
  ``Perturbations around backgrounds with one non-homogeneous dimension,''
  arXiv:hep-th/0609001.

\bibitem{power}
  B.~Kol,
  ``The power of action: 'The' derivation of the black hole negative mode,''
  arXiv:hep-th/0608001.

\bibitem{rev}
  B.~Kol,
  ``The phase transition between caged black holes and black strings: A
  review,''
  Phys.\ Rept.\  {\bf 422}, 119 (2006)
  [arXiv:hep-th/0411240].

\bibitem{HOrev}
  T.~Harmark, V.~Niarchos and N.~A.~Obers,
  ``Instabilities of black strings and branes,''
  arXiv:hep-th/0701022.

\bibitem{TopChange}
  B.~Kol,
 ``Topology change in general relativity and the black-hole
 black-string transition,''
  JHEP {\bf 0510}, 049 (2005)
  [arXiv:hep-th/0206220].

\bibitem{RW}
T.~Regge and J.A.~Wheeler, ``Stability of a Schwarzschild
  singularity'', \pr{108}{1957}{1063}.

\bibitem{Zerilli}
  F.~J.~Zerilli,
  ``Effective Potential For Even Parity Regge-Wheeler Gravitational
  Perturbation Equations,''
  Phys.\ Rev.\ Lett.\  {\bf 24}, 737 (1970).

\bibitem{IshibashiKodama}
  H.~Kodama and A.~Ishibashi,
  ``A master equation for gravitational perturbations of maximally  symmetric
  black holes in higher dimensions,''
  Prog.\ Theor.\ Phys.\  {\bf 110}, 701 (2003)
  [arXiv:hep-th/0305147].

\bibitem{dialogue1}
  D.~Gorbonos and B.~Kol,
  ``A dialogue of multipoles: Matched asymptotic expansion for caged black
  holes,''
  JHEP {\bf 0406}, 053 (2004)
  [arXiv:hep-th/0406002].

\bibitem{IshibashiKodama2}
  A.~Ishibashi and H.~Kodama,
  ``Stability of higher-dimensional Schwarzschild black holes,''
  Prog.\ Theor.\ Phys.\  {\bf 110}, 901 (2003)
  [arXiv:hep-th/0305185].

\bibitem{Moncrief1974}
  V.~Moncrief,
   ``Gravitational perturbations of spherically symmetric systems. I. The
  exterior problem,''
  Annals Phys.\  {\bf 88}, 323 (1974).

\bibitem{MartelPoisson}
  K.~Martel and E.~Poisson,
   ``Gravitational perturbations of the Schwarzschild spacetime: A practical
  covariant and gauge-invariant formalism,''
  Phys.\ Rev.\ D {\bf 71}, 104003 (2005)
  [arXiv:gr-qc/0502028].

\bibitem{NagarRezzolla}
  A.~Nagar and L.~Rezzolla,
   ``Gauge-invariant non-spherical metric perturbations of Schwarzschild
  black-hole spacetimes,''
  Class.\ Quant.\ Grav.\  {\bf 22}, R167 (2005)
  [Erratum-ibid.\  {\bf 23}, 4297 (2006)]
  [arXiv:gr-qc/0502064].

\bibitem{AKS}
  V.~Asnin, B.~Kol and M.~Smolkin,
  ``Analytic evidence for continuous self similarity of the critical merger
  solution,''
  Class.\ Quant.\ Grav.\  {\bf 23}, 6805 (2006)
  [arXiv:hep-th/0607129].

\bibitem{nSchw}
  B.~Kol,
  ``A new action-derived form of the black hole metric,''
  arXiv:gr-qc/0608001.

\bibitem{Higuchi:1986wu}
  A.~Higuchi,
  ``Symmetric tensor harmonics on the N sphere and their application
  to the de Sitter group SO(N,1),''
  J.\ Math.\ Phys.\  {\bf 28}, 1553 (1987)
  [Erratum-ibid.\  {\bf 43}, 6385 (2002)].
\end{thebibliography}
\end{document}